# Electron Doping Stabilization of Highly-Polar Supertetragonal BaSnO$_3$

Qing Zhang[1], Karin M. Rabe[2] and Xiaohui Liu[1,*]

[1] *School of Physics, Shandong University, Ji'nan 250100, China*

[2] *Department of Physics and Astronomy, Rutgers University, Piscataway, New Jersey 08854, USA*

Could electrons stabilize ferroelectric polarization in unpolarized system? Basically, electron doping was thought to be contrary to polarization due to the well-known picture that the screening effect on Coulomb interaction diminishes ferroelectric polarization. However, in this paper, we propose a novel mechanism of stabilizing highly-polar supertetragonal BaSnO$_3$ by electron doping. With moderate compressive strain applied, less than -5.5%, BaSnO$_3$ exhibits stable nonpolarized normal tetragonal structure and an unstable supertetragonal state which is characterized with extremely large c/a ratio and giant polarization. We found that the band gap of the supertetragonal state is much smaller than the normal tetragonal state, with a difference around 1.2eV. Therefore, the energy of the doped electrons selectively favors the smaller gap supertetragonal state than the larger band gap normal tetragonal state, and the critical strain to stabilize the supertetragonal phase could be reduced by electron doping. This mechanism guarantees the controllable supertetragonal structures by electron doping and ensures the coexistence of giant polarization and conducting in high-mobility BaSnO$_3$, and is promising to design high-mobility ferroelectrics conductor.

**Introduction**

The understanding of ferroelectricity in perovskite oxides ABO$_3$ centers on the identification of lattice instabilities of the cubic perovskite reference structure that can be frozen in to generate electric-field-switchable polar phases. The prototypical cases are PbTiO$_3$ and BaTiO$_3$, for which the cubic structure has an unstable zone-center polar mode that freezes in and couples to strain to generate the P4mm tetragonal phase, with c/a of 1.06 and 1.01 respectively, and the additional orthorhombic and R3m phases of BaTiO$_3$ [1, 2].

In a number of perovskite oxides, it has emerged that there is a distinct highly-polar tetragonal P4mm phase, related to the known ferroelectric phases by breaking of one of the B-O bonds along c and dramatic elongation of the unit cell to give c/a ratios on the order of 1.2 [3, 4]. This "supertetragonal" (ST) phase has been observed as the equilibrium structure only in PbVO$_3$ and BiCoO$_3$ [5,6]. In PbTiO$_3$, first-principles calculations showed that a locally stable supertetragonal structure could be stabilized with negative pressure, and in BaTiO$_3$ with a much higher negative pressure [7]. The negative pressure induced ST PbTiO$_3$ was observed in nanowire [8]. Another way to realized ST PbTiO$_3$ is reported by mixed growth of PbTiO$_3$ and PbO, where PbO provides longitudinal strain on PbTiO$_3$ [9]. In BiFeO$_3$, the ST phase can be straightforwardly stabilized by compressive epitaxial strain [10,11] at an extremely large compressive strain at least -4.5% is required (matching the c/a change) – can be realized by an available substrate.

BaSnO$_3$ is an insulating cubic perovskite oxide which has recently been of great interest as a parent compound to design transparent conducting oxides of high mobility by electron doping due to its isolated Sn s conduction band [12,13]. The in-plane epitaxial compressive strain $\eta$ (less than -5.5%) tunes BaSnO$_3$ from cubic structure to unpolarized tetragonal P4/mmm structure ("normal tetragonal" NT). Besides the stable NT structure, first-principles calculations show a competing polarized ST state which can be stabilized by compressive epitaxial strain greater than a critical value around -5.5% [14, 15]. Growth on an SrTiO$_3$ (001) substrate however appears not to provide a sufficiently high value for the compressive strain for observation of the ST phase.

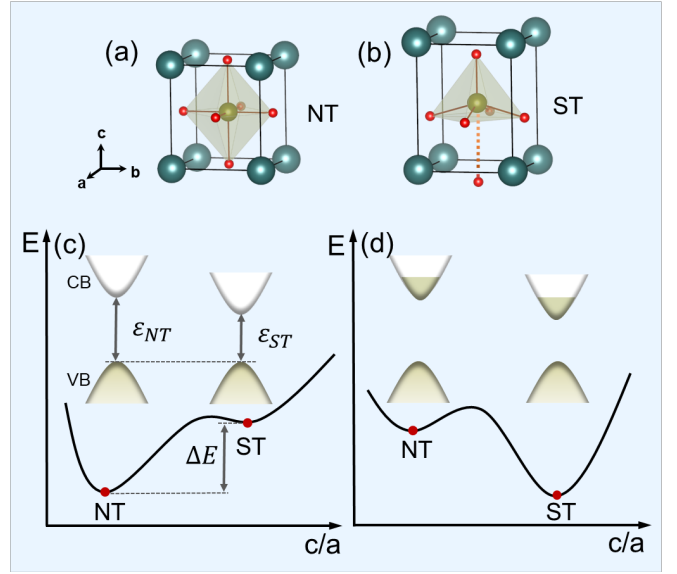

Fig. 1. (a) Normal tetragonal (NT) and (b) supertetragonal (ST) perovskite structure. (c) Energy profile of BaSnO$_3$ as a function of c/a ratio without doping. The difference of energy between the two states is $\Delta E$. The schematic plots of the conduction band (CB) and valence band (VB) of NT and ST states indicate the different band gaps, $\varepsilon_{NT} > \varepsilon_{ST}$. (d) Energy profile of BaSnO$_3$ with electron doping $n$, when $n(\varepsilon_{NT} - \varepsilon_{ST}) > \Delta E$.

In this paper, we focused on BaSnO$_3$ to explore a mechanism to reduce the value of the critical epitaxial strain for stabilization of the ST structure. Our strategy, illustrated in Fig.1, stands on the fact that the polarized ST state of BaSnO$_3$ has a much smaller band gap $\varepsilon_{ST}$ than $\varepsilon_{NT}$ the band gap of the unpolarized NT state. The doping of electrons into BaSnO$_3$ thus favors the ST structure of BaSnO$_3$ over the NT structure, by the smaller additional electronic energy of the doped electrons. To be specific, as shown in Fig. 1(c), if for undoped BaSnO$_3$ the epitaxial strain dependence of the energy difference between NT and ST states per 5-atom unit cell is $\Delta E(\eta)$ (the critical value is at $\Delta E = 0$ ). With doping concentration $n$, the additional electronic energy of the NT state is larger than the ST state, estimated as $\Delta \varepsilon = n(\varepsilon_{NT} - \varepsilon_{ST})$. When the doping concentration is above a critical value $n_0(\eta)$, then $\Delta \varepsilon > \Delta E$, and the structure undergoes a first order phase transition from the large gap NT state to the smaller gap ST state to lower its energy, as shown in Fig. 1(d). In the following, we perform calculations to quantify this mechanism in strained BaSnO$_3$, and show that reasonable doping levels substantially lower the critical epitaxial strain for the formation of the ST phase.

**Methods**

First-principles density functional theory (DFT) calculations are performed by using Vienna Ab initio Simulation Package (VASP). For calculations with 5-atom unit cells, the hybrid functional HSE06 [16] and projector augmented wave methods are used with an energy cutoff of 500 eV. The Brillouin zone is sampled with a 6 × 6 × 6 Monkhorst-Pack k-point meshes for structural relaxation with force convergence criterion less than 10 meV/Å. For supercell calculations, we use the GGA functional [17] with an energy cutoff of 500 eV. The Brillouin zone is sampled with 3 × 3 × 3 and 4 × 4 × 4 Monkhorst-Pack k-point meshes for supercells built by 3×3×3 and 2×2×2 primitive cell, respectively. In Supplementary Materials Table S1, we provide the calculated lattice parameters of cubic BaSnO$_3$ using different functionals, as well as experimental lattice parameters. The lattice parameter 4.132 Å, calculated using HSE06 hybrid functional, matches well with the experimental value. The electron doping is achieved by adding extra electrons to the system with the same amount of uniform positive charge in the background.

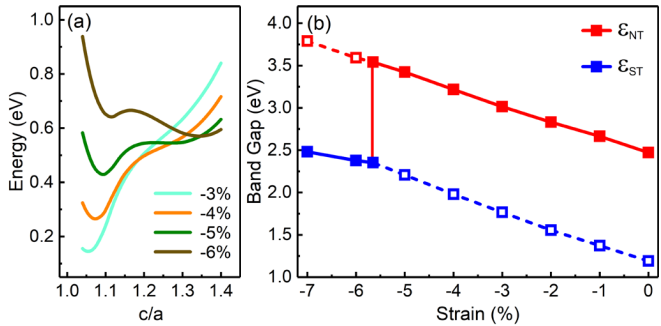

Fig. 2. (a) Energy profiles as a function of c/a ratio for four fixed in-plane strain values. The energy of cubic BaSnO$_3$ is taken as the zero of energy. (b) Band gaps as a function of the compressive strain for the NT state (red) and ST state (blue). The solid symbols indicate the band gap of the ground state, and the open symbols indicate the unstable states. The solid lines and dashed lines indicate the band gap of the ground state and the unstable higher energy state, respectively.

**Results**

In Fig. 2(a), we show the energy of the relaxed structure of P4mm BaSnO$_3$ as a function of fixed c/a ratio for compressive epitaxial strains ranging from -3% to -6%. From the shape of the curve for each epitaxial strain, we see that there are two local minima, one at low and the other at high c/a, corresponding to two competing states: the unpolarized NT state and the polarized ST state, respectively. When the compressive strain is less than about -5%, the unpolarized NT state is the ground state while the polarized ST state is at higher energy, as illustrated by the energy profile in Fig. 1(c). Detailed calculations show that the polarized ST state becomes the ground state when the compressive strain is larger than a critical value of around -5.5%. For example, at compressive strain of -6% strain as shown by Fig. 2(a), the energy profile corresponds to that in Fig. 1(d).

The band gaps $\varepsilon_{ST}$ and $\varepsilon_{NT}$ of the two states as a function of the compressive strain are shown in Fig. 2(b). The band gaps of both states are almost linearly dependent on strain with nearly the same slope, so that the difference in band gap is almost energy independent: from 0% to 7% strain, the band gap of the ST state is smaller than the band gap of NT state by about 1.2 eV, corresponding to the band extrema shown in Fig. 1(c) and (d). Further discussion is provided in the Supplementary Materials.

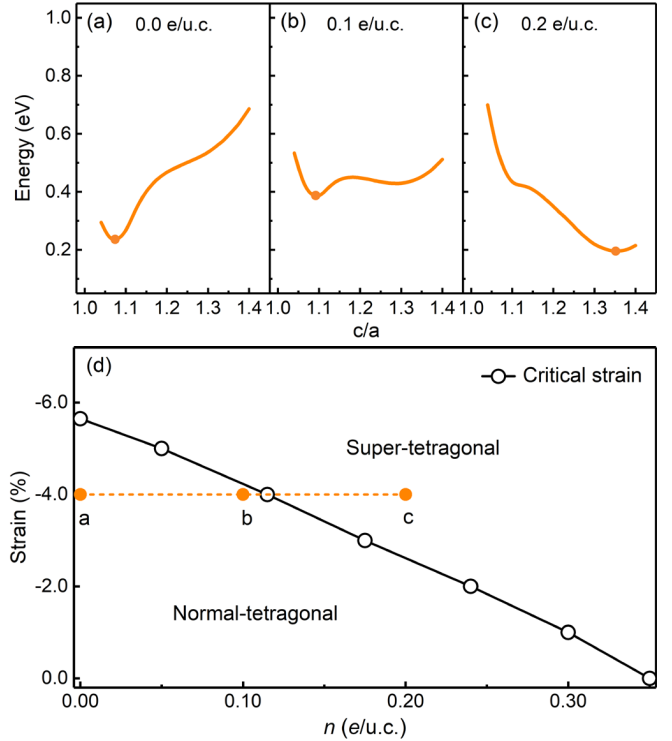

Fig. 3. Energy profile as a function of c/a ratio for in-plane strain -4% with different doping concentration: (a) without doping, (b) 0.1 e/u.c. and (c) 0.2 e/u.c.. (d) doping-strain phase diagram of NT and ST BaSnO$_3$. The open circles correspond to the calculated critical strain with different doping concentration, and the solid curve indicates the boundary between the two phases. The corresponding positions of the stable state in (a), (b) and (c) is signed by the points in the phase diagram by the orange points.

Next, we show that electron doping favors the ST state. Here, we take epitaxial in-plane strain to be -4% to illustrate the principle. For other strains, the results are given in the Supplementary Materials. Energy profiles as a function of c/a ratio for different doping concentrations are shown in Fig. 3. Without doping, the energy profile Fig. 3(a) which is the same as the orange curve of -4% in Fig. 2(a), the NT state with c/a ratio about 1.07 is stable while the ST state is unstable. With doping concentration 0.1e/u.c. as shown in Fig. 3(b), the energy of the ST state is lowered relative to that of the NT state, though the latter is still the ground state. With the doping concentration increased to 0.2e/u.c. (Fig. 3(c)), the ST state, with c/a ratio about 1.35, is the ground state.

For each strain, we calculated the minimum doping concentration $n_0(\eta)$ required to stabilize the NT state. The resulting doping-strain phase diagram, with the phase boundary separating NT and ST states, is shown in Fig. 3(d). It is clearly to see that the critical strain to stabilize the ST state decreases as a function of the doping concentration. In particular, with doping concentration 0.12e/u.c., the critical compressive strain required to stabilize ST is -4%, which is much smaller than the critical strain of -5.5% to stabilize the ST state without doping and well below the mismatch strain for growth on SrTiO$_3$ (001).

Next, we confirm that the lower energy of the ST state with doping arises from the lower energy of added electrons, as illustrated by the mechanism in Fig. 1(c) and (d). In Fig. 4, we show the energy profiles $E_{ST}$ of the undoped ST state and $E_{NT}$ of the undoped NT state as a function of compressive strain. The energies are equal at around -5.5% which is the critical strain without doping for this first-order transition. Below the critical strain, $E_{ST}$ is higher than $E_{NT}$, and the energy difference of the two states $\Delta E = (E_{ST} - E_{NT})$ is shown. For each value of strain, we obtained the critical doping concentration $n_0(\eta)$ from the strain-doping phase diagram and compute $n_0(\varepsilon_{NT} - \varepsilon_{ST})$, the reduction in one-electron energy due to the reduction of band gap from NT state to ST state. We see that $n_0(\varepsilon_{NT} - \varepsilon_{ST})$ accounts for most of $\Delta E$ for all strains considered, confirming the mechanism in Fig. 1(c) and (d).

In the calculations above, the net charge of the added electrons was artificially compensated by a uniform positive background, as explained in the Methods section. In real systems, the electrons are added into BaSnO$_3$ by substitution of A-site cation with La, substitution of B-site cation with Nb. Here, we extend our calculations to consider the effects of realistic electron doping on the stability of the ST state by using supercells to describe systems with substitutions. In these calculations, due to the extreme computational demands of HSE06, we use GGA. As a reference, we re-performed the computations for uniform positive background with GGA. It is well known that the band gap given within GGA is underestimated, and the difference of band gaps between NT and ST states is around 0.80 eV which is smaller than the value 1.2 eV given by HSE06, decreasing the energy gain of the ST phase with doping. On the other hand, GGA tends to overestimate unit-cell volume, which acts like a negative pressure to favor the ST state. In fact, the calculated critical strain given by GGA is around -5.25% which is very close to -5.5% given by HSE06, and we use this to inform our analysis of the GGA supercell results.

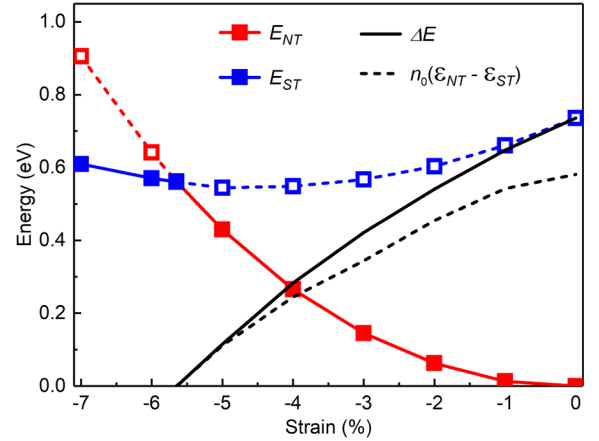

Fig.4 Energy profiles of the undoped BaSnO$_3$ of the NT (red) and ST (blue) as a function of compressive strain, and the difference between them is plotted by the solid black curve; The dashed black curve corresponds to the reduced electronic energy with minimum electron doping $n_0$ required to stabilize the ST state. The solid of red and blue mark the ground state.

In Fig. 5, we show the critical strain as a function of the doping concentration by substitution of Ba with La, substitution of Sn with Nb. The critical strain is always less than -5.25%, and the critical strain decreases when the doping concentration is increased which matches the results talked above.

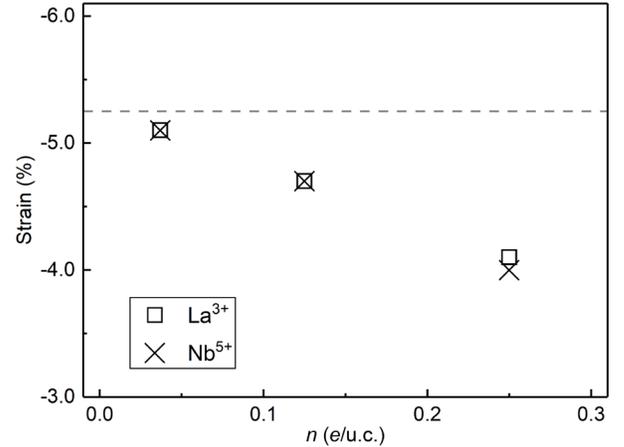

Fig.5 Critical strain as a function of electrons doping with different doping methods: A-site substation with La, B-site substitution with Nb. The dashed line indicates the critical strain without doping.

**Discussion**

Our first-principles calculations provide strong evidence that electron doping of BaSnO$_3$ can substantially lower the critical value of epitaxial strain needed to stabilize the supertetragonal phase via the physical mechanism illustrated in Fig. 1. In experiment, for example, it was reported that coherent growth of an ultra-thin layer of BaSnO$_3$ on SrTiO$_3$ could be realized, with a compressive strain of around -5% from the lattice mismatch [18]. Though the observed phase was NT for the undoped film, our calculations suggest that modest doping levels could lower the critical value of strain below this value, allowing the epitaxial growth of the ST phase. Another promising experimental configuration could be heterostructures in which the $BaSnO_3$ layer is modulation doped.

We mention that the applied strain may induce the rotation of the oxygen octahedra. As described in the Supplementary Materials, we have performed calculations for the systems with uniformly charged background to find that octahedron rotation doesn't have significant effects on the critical strain and band gap which are the main points of this work. The supercell calculations include allowed oxygen octahedron rotation distortions, and we see that the critical strains are not increased by their presence.

This approach to stabilizing the ST phase has an intriguing advantage – electron-doped ST $BaSnO_3$ is predicted to be a ferroelectric metal [19, 20], a material class of great current interest [21, 22, 23, 24]. Moreover, the mobility of the added electrons is expected to be very high, based on the high mobility of added electrons in cubic $BaSnO_3$ [12, 13] and the fact that the isolated low-mass character of the Sn conduction band minimum is essentially the same in the ST and NT phases. Ferroelectric materials with high mobility are not only interesting for fundamental interests but also promising for practical applications such as solar cells, ferroelectric field effects transistors, and photovoltaic devices from the aspect practical applications.

More generally, it is increasingly established that many perovskite oxides have competing ST states. Here, in addition to the stabilization by compressive epitaxial strain or negative pressure, we have another control parameter for perovskite oxides in which the band gap for the ST phase is significantly smaller than in the NT phase. For example, as shown in the Supplementary Materials, the ground state of $SrSnO_3$ undergoes paraelectric Pbnm and I4/mcm structure with compressive applied, while its ST state always has a smaller band gap by about 1.5 eV. This means that the electron doping in $SrSnO_3$ could also assist the stabilization of the ST phase.

**Summary**

In conclusion, we studied the effects of electron doping on the reduction of critical strain to stabilize the highly-polar ST phase of $BaSnO_3$. The energy of the added electrons selectively favors the ST phase over the NT phase, which has a larger band gap. The predicted effect provides a novel mechanism to design conducting ferroelectrics with giant polarization in perovskite oxides.

**Acknowledgement**

The work of X. L. was supported by the National Natural Science Foundation of the People's Republic of China (Grants 11974211) and Qilu Young Scholar Program of Shandong University. The work of K. M. R. was supported by the Office of Naval Research through grant N00014-21-1-2107. We acknowledge useful discussions with Charles Ahn, Frederick Walker from Yale University, and Zheng Wen from Qingdao University.

liuxiaohui@sdu.edu.cn

Supplementary Materials

**Electron Doping Stabilization of Highly-Polar Supertetragonal BaSnO$_3$**

Qing Zhang[1], Karin M. Rabe[2] and Xiaohui Liu[1]

[1] *School of Physics, Shandong University, Ji'nan 250100, China*

[2] *Department of Physics and Astronomy, Rutgers University, Piscataway, New Jersey 08854, USA*

## A. Lattice parameters of cubic BaSnO$_3$ and SrTiO$_3$

Table S1. The calculated lattice parameters of cubic BaSnO$_3$ and SrTiO$_3$ using different DFT functionals, as well as reported experimental values.

|  | DFT functionals | | | | Exp [1-2] |
| --- | --- | --- | --- | --- | --- |
|  | LDA | PBE | PBEsol | HSE06 |  |
| BaSnO$_3$  $a$ (Å) | 4.098 | 4.187 | 4.134 | 4.132 | 4.116 |
| SrTiO$_3$  $a$ (Å) | 3.861 | 3.940 | 3.895 | 3.898 | 3.905 |

We calculated the lattice parameters of cubic BaSnO$_3$ and SrTiO$_3$ using different functionals. For both BaSnO$_3$ and SrTiO$_3$, we found that the lattice parameters calculated using hybrid functional HSE06 are more close to the experimental values.

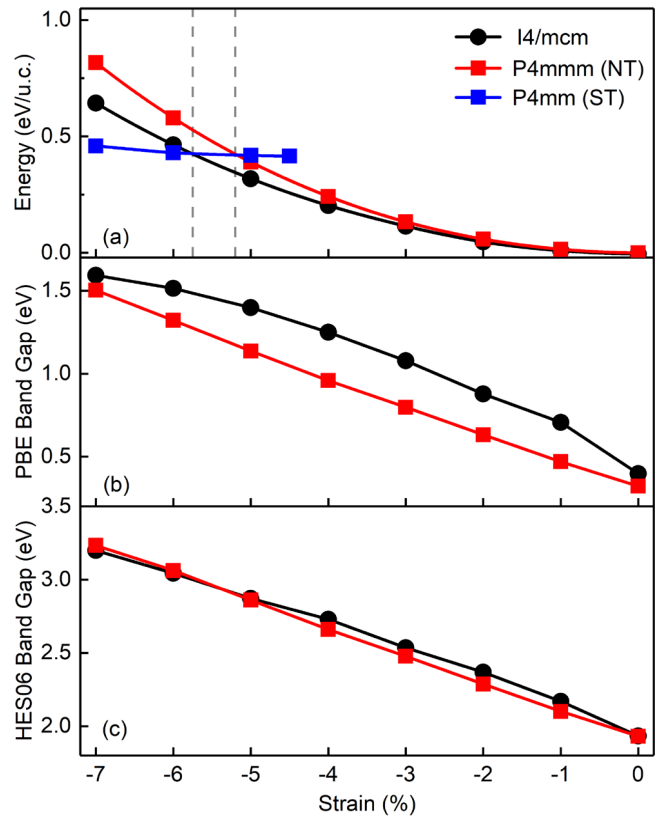

Fig. 1S. (a) Energy profiles as a function of compressive strain for different phases of BaSnO$_3$. Band gaps as a function of compressive strain for the P4mmm and I4/mcm states calculated using different functionals: (b) GGA and (c) HSE06.

The compressive strain may lead to the emergence of oxygen octahedral rotation in BaSnO$_3$. We found that, with compressive strain being applied, I4/mcm structure is more stable. Fig. 1S shows the energy profiles of the various phases as a function of strain for BaSnO$_3$. When octahedral rotation is considered, the critical strain has a small variation of around 0.4% as shown by Fig. 1S(a). Also, we found that the octahedral rotation doesn't have significant effects on the band gaps, as shown by Fig. 1S (b) and (c). The I4/mcm structure even has larger band gaps than the normal tetragonal P4mmm structure. It means that, compared with the normal tetragonal P4mmm structure, the I4/mcm structure is more inclined to transfer to the supertetragonal P4mm structure when electrons are doped.

**B. Band Alignment between NT and ST structures.**

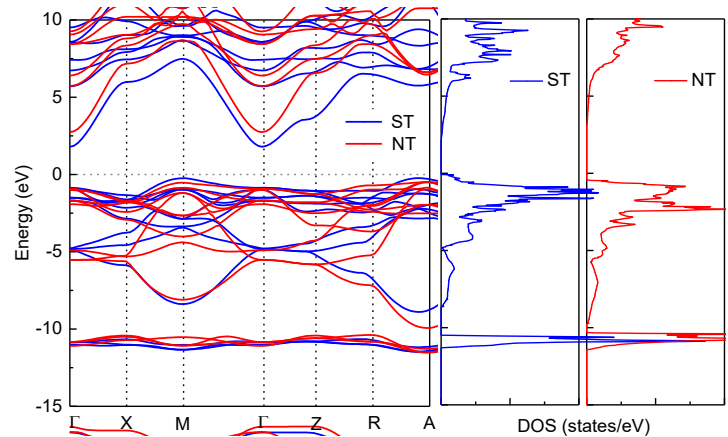

Fig. 2S. Band alignment between the ST and NT structures with compressive -4% applied.

In Fig. 2S, we calculated the bands of the ST and NT structures of BaSnO$_3$. The Ba semi-core 5p state is lined up. The density of states of the two structures are also plotted. It is clearly to see that the conduction band of ST structure is lower than the conduction band of NT structure around 1.2 eV.

**C. The effects of electron doping on the stability of the ST state**

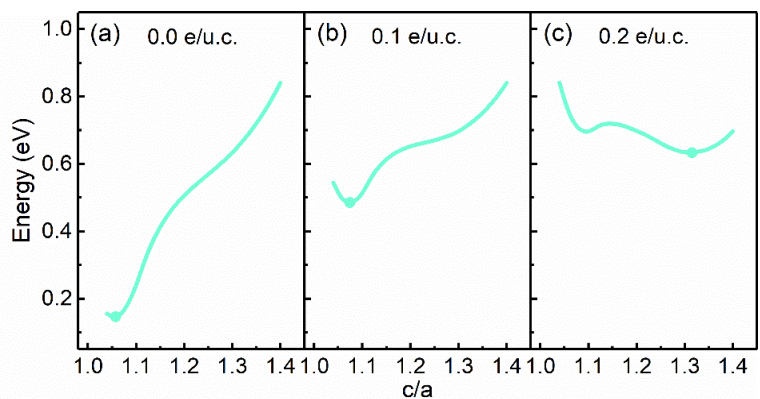

Fig. 3S. Energy profiles as a function of c/a ratio for in-plane strain -3% with doping concentration: (a) without doping, (b) 0.1 e/u.c. and (c) 0.2 e/u.c..

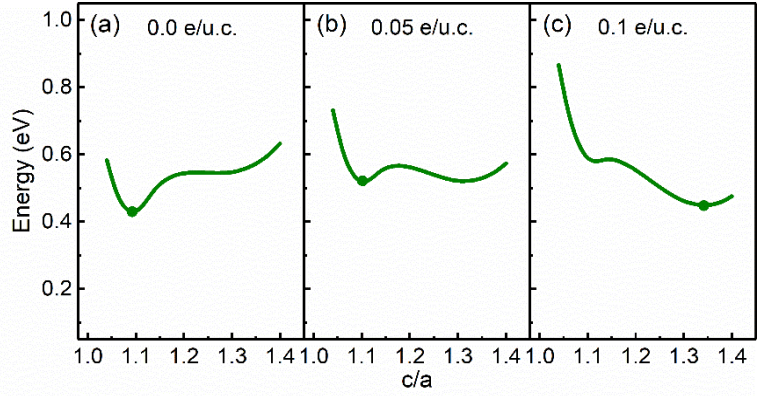

Fig. 4S. Energy profiles as a function of c/a ratio for in-plane strain -5% with doping concentration: (a) without doping, (b) 0.05 e/u.c. and (c) 0.1 e/u.c..

In addition to the effects of electron doping on the stability of the ST state for in-plane strain -4% shown in Fig. 3 in the main text, we also examined the cases of -3% and -5% strain shown in Fig.2 (a), as shown in Fig. 3S and Fig. 4S, respectively. Similarly, we observe that the ground state is NT without doping. As the doping concentration increases, the difference of the energy between the ST and NT states decreases. With the doping concentration increased to 0.2e/u.c. (0.1e/u.c.), the ground state is reversed to ST state for in-plane strain -3% (-5%).

## D. $SrSnO_3$

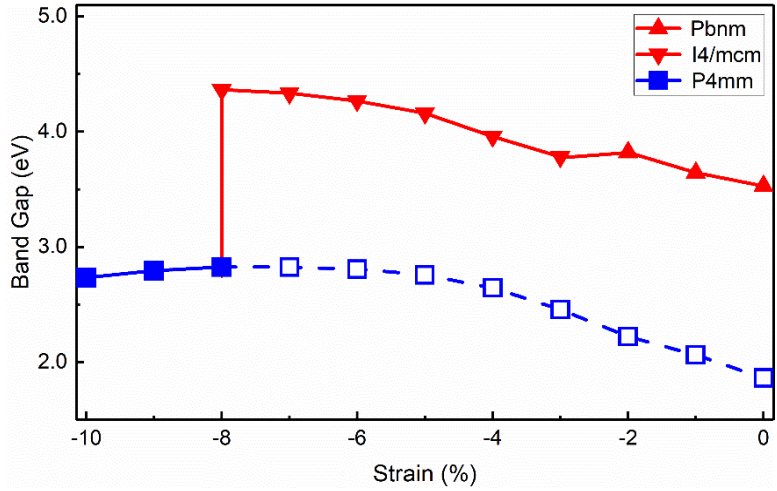

Fig. 5S. Band gaps as a function of the compressive strain for the ST state (blue) and ground states (red) of $SrSnO_3$.

The ground state of $SrSnO_3$ is Pbnm structure. When compressive is applied, it goes to I4/mcm structure at strain about -2.0%, and then transits into supertetragonal P4mm structure. As we shown in Fig 5S, the supertetraonal P4mm structure always has much smaller band gap by around 1.5 eV.

[1] T. Maekama, K. Kurosaki, S. Yamanaka, J. Alloys Compd. **416**, 214 (2006).
[2] Janotti, A., Jalan, B., Stemmer, S. and Van de Walle, C. G., Appl. Phys. Lett., **100**, 26 (2012).